\documentstyle[12pt,epsfig]{article}

\newcommand{\be}{\begin{eqnarray}}
\newcommand{\ee}{\end{eqnarray}}

\def\nue{{\nu_e}}

\def\numu{{\nu_{\mu}}}

\def\nutau{{\nu_{\tau}}}

\newcommand{\dm}{\mbox{$\Delta m_{21}^2$~}}

\newcommand{\br}{\mbox{$^{8}{B}~$}}
\newcommand{\ber}{\mbox{$^{7}{Be}$~}}

\newcommand{\kl}{\mbox{KamLAND~}}
\newcommand{\bx}{\mbox{Borexino~}}
\newcommand{\thsol}{\mbox{$\theta_{12}$~}}
\def\lsim{\:\raisebox{-0.5ex}{$\stackrel{\textstyle<}{\sim}$}\:}
\def\gsim{\:\raisebox{-0.5ex}{$\stackrel{\textstyle>}{\sim}$}\:}
\begin{document}

\begin{flushright}
SISSA 15/2003/EP     \\
SINP/TNP/03-04\\
hep-ph/0302243
\end{flushright}

\begin{center}
{\Large \bf Exploring the sensitivity of current and future 
experiments to $\theta_{\odot}$}
\vspace{.5in}

{\bf Abhijit Bandyopadhyay\footnote{e-mail: abhi@theory.saha.ernet.in},
 Sandhya Choubey\footnote{email: sandhya@he.sissa.it},
 Srubabati Goswami\footnote{e-mail: sruba@mri.ernet.in}}
\vskip .5cm

$^1${\it Theory Group, Saha Institute of Nuclear Physics},\\
{\it 1/AF, Bidhannagar,
Calcutta 700 064, INDIA}\\
$^2${\it INFN, Sezione di Trieste and
Scuola Internazionale Superiore di Studi Avanzati,\\
I-34014,
Trieste, Italy}\\
$^3${\it Harish-Chandra Research Institute},\\{\it Chhatnag Road, Jhusi,
Allahabad  211 019, INDIA}

\vskip 1in

\end{center}

\begin{abstract}
The first results from the KamLAND experiment 
in conjunction with the global solar neutrino data
has demonstrated striking ability to 
constrain the $\Delta m^2_\odot$ ($\Delta m^2_{21}$) very precisely.
However the allowed range of 
$\theta_{\odot}$ ($\theta_{12}$) did not change much with the inclusion of the 
KamLAND results. 
In this paper we probe if future data from KamLAND can increase the accuracy 
of the allowed range in $\theta_{\odot}$ and conclude that even after 
3 kton-year of statistics 
and most optimistic error estimates, 
KamLAND {\it may} find it hard to 
significantly improve the 
bounds on the mixing angle obtained from the solar neutrino data. 
We discuss the $\theta_{12}$ sensitivity of the survival probabilities in 
matter (vacuum)  as is 
relevant for the 
solar (KamLAND)  experiments. 
We find that the presence of matter effects in the survival probabilities
for $^8B$ neutrinos give the solar neutrino experiments
SK and SNO an edge over KamLAND,
as far as $\theta_{12}$ 
sensitivity is concerned, particularly near maximal mixing.
Among solar neutrino experiments we identify SNO as the most 
promising candidate for constraining $\theta_{12}$ and make a 
projected sensitivity test for the mixing angle by reducing the 
error in the neutral current measurement at SNO. Finally we 
argue that the most accurate bounds on $\theta_{12}$ can be  
achieved in a reactor experiment, 
if the corresponding baseline and energy 
can be tuned to a minimum in the survival probability. 
We propose a new reactor experiment which can give 
the value of $\tan^2\theta_{12}$ to 
within 14\%.
We also discuss the future Borexino and LowNu experiments. 
 
\end{abstract}

\newpage

\section{Introduction}

The year 2002 has witnessed two very important results in solar
neutrino research.
In April 2002 the accumulated evidence in
favor of  possible flavor conversion
of the solar electron neutrinos was confirmed with a statistical
significance of 5.3$\sigma$ from the 
Sudbury Neutrino Observatory (SNO) 
\cite{Ahmad:2002jz}.
The inclusion of the SNO spectrum data combining the charged current,
electron scattering and neutral current events in the global solar neutrino
analysis picked out
the Large Mixing Angle (LMA) MSW \cite{msw} solution as the preferred
solution \cite{Bandyopadhyay:2002xj,Choubey:2002nc},
confirming the earlier  trend \cite{snoccgl}.
In December 2002 the Kamioka Liquid scintillator Anti-Neutrino 
Detector (KamLAND) experiment in Japan \cite{Eguchi:2002dm}
provided independent and conclusive 
evidence in favor of the LMA solution, using
reactor neutrinos.
Assuming CPT invariance this establishes oscillations of $\nu_e$
with a mass squared difference $\sim 7 \times 10^{-5}$ eV$^2$ 
and large mixing \cite{Bandyopadhyay:2002en,others}.
Comprehensive evidence in favor
of oscillation of the atmospheric $\nu_{\mu}$s came from the
Super-Kamiokande (SK) 
results \cite{skatm}. This was confirmed by the result from
the K2K long baseline experiment using terrestrial neutrino sources \cite{k2k}.
The best-fit value of $\Delta m^2_{atm}$ comes out as $\sim 2.5 \times 10^{-3}$
eV$^2$ with maximal mixing in the $\numu-\nutau$ sector \cite{kajita}.

Since the solar and atmospheric neutrino anomalies involve two hierarchically
different mass scales, simultaneous explanation of these involve three
neutrino mixing. There are nine unknown parameters involved in the 
three-generation light neutrino mass matrix -- masses of the three 
neutrinos, and six other parameters coming from the 
Pontecorvo-Maki-Nakagawa-Sakata (PMNS) mixing matrix \cite{pmns}.
Of the nine parameters, 
oscillation experiments are sensitive to six ($\Delta m^2_{21}$,
$\Delta m^2_{31}$, $\theta_{12}$, $\theta_{13}$, $\theta_{23}$, $\delta$),
the two independent 
mass squared differences ($\Delta m_{ij}^2=m_i^2 - m_j^2$), 
the three mixing angles and one CP 
phase. 
Flavor oscillations are independent of 
the absolute neutrino mass scale, and 
the remaining two CP phases appear only in lepton number violating 
processes. 
The solar neutrino data constrain the parameters $\Delta m^2_{\odot}
\sim  \Delta m^2_{21}$ and $\theta_{\odot} \sim \theta_{12}$ while
the atmospheric neutrino data constrain the parameters
$\Delta m^2_{atm} \sim \Delta m_{31}^2$ and $\theta_{atm} \sim \theta_{23}$.
The two sectors get connected by the mixing angle $\theta_{13}$ which
is at present constrained by the reactor data \cite{chooz,Boehm:2001ik}
as $\sin^2 \theta_{13} \leq 0.03$ at 90\% C.L. \cite{chooz}.

With neutrino flavor oscillations in 
both solar and atmospheric neutrino anomalies confirmed, the
research in neutrino physics is now all set to enter the
era of precision measurements.
The conventional accelerator based
long baseline experiments as well as neutrino factories using muon storage
rings as sources have been
discussed widely for the purposes of precise determination of the 
neutrino oscillation parameters
(see \cite{nufact} for a comprehensive discussion and 
a complete list of references). 
The major goals in the upcoming long baseline and proposed neutrino factories
are -- precision determination of $|\Delta m^2_{31}|$ and
$\theta_{23}$ , ascertaining the sign of $\Delta m^2_{31}$ and
determining how small is $\theta_{13}$.
The atmospheric parameters 
$|\Delta m_{31}^2|$ and $\sin^22\theta_{23}$ are expected to be 
determined within 1\% accuracy in the next generation long baseline 
experiments using conventional 
(super)beams \cite{Itow:2001ee,Ayres:2002nm}. The 
mixing angle $\sin^22\theta_{13}$ is expected to be probed down 
to $1.5\times 10^{-3}$ in the long baseline experiments using 
superbeams \cite{Itow:2001ee,Ayres:2002nm}
while neutrino factories will be sensitive upto  
$\sin^22\theta_{13} \sim 10^{-5}$ \cite{nufact}.
Finally with KamLAND confirming the LMA solution,
it should be possible to
measure the CP phase $\delta$ in neutrino factories 
and possibly even in the proposed phase II JHF (in Japan) and NuMI (in USA) 
long baseline experiments,
provided $\sin^22\theta_{13}$ is not too small \cite{Itow:2001ee,Ayres:2002nm}.
However $\Delta m^2_{21}$ and $\theta_{12}$ drive the sub-leading oscillations
in these experiments and hence precision determination of these parameters
through long baseline experiments or neutrino factories will be very
challenging\footnote{With LMA confirmed by KamLAND 
there remains a possibility of determining $\theta_{12}$ and 
$\Delta m^2_{21}$ in a high-performance neutrino factory provided 
the background can be reduced sufficiently and
$\sin^2 2\theta_{13} < 10^{-5}$ \cite{nufact21}.}. Therefore in 
all these studies the sub-leading oscillation parameters $\Delta m^2_{21}$
and $\theta_{12}$ are introduced as external inputs, taking typically 
either the
best-fit value obtained from the global solar analysis or the projected 
sensitivity limits from future \kl data.
However, since the concern now has shifted to precision measurements,
the uncertainty in the parameters $\Delta m^2_{21}$
and $\theta_{12}$
can also affect the accuracy with which we can determine the rest of the
parameters of the PMNS matrix, 
especially the CP violation parameter $\delta$, as
it comes only with the sub-leading term in the oscillation probability.
The uncertainty in the measurement of other parameters, 
introduced through the uncertainty in the solar parameters, gets 
worse for smaller values of the mixing angle $\sin^22\theta_{13}$.

As far as the precision determination of $\Delta m^2_{21}$ is concerned,
KamLAND
has already demonstrated an extraordinary capability in precisely determining
the $\Delta m^2_{\odot}$.  The 
uncertainty (we call it ``spread'')\footnote{We give the precise
definition of ``spread'' in Section 3.} in the 99\% C.L.
allowed range of this parameter around the global best-fit solution 
(which we call the low-LMA),
has reduced to 30\% after including the KamLAND spectral data,
from 76\% as obtained from only solar global analysis.
The spread in the allowed range of $\tan^2\theta_{\odot}$ on the other hand
remains
unchanged, even after including the KamLAND results and the current 99\% C.L.
uncertainty is $\sim$ 47\%.

In this paper we probe the sensitivity of the various previous, 
present and future solar neutrino
experiments to the parameter $\theta_{\odot} \sim \theta_{12}$ and make a
comparative study of which experiment is most sensitive in constraining
$\theta_{12}$. We conclude that SNO has the best potential for constraining 
\thsol. We make an optimistic projected analysis including 
future SNO neutral current (NC) measurements and look for the improved 
bounds on \thsol. 
We discuss the precipitating factors
for which the sensitivity of KamLAND to $\theta_{12}$ is not
as good as its sensitivity to $\Delta m^2_{21}$ and discuss
the effect of increased statistics and reduced systematics through projected
analyses. We conclude that even with 3 kTy statistics \kl may
not significantly improve the current limits on \thsol coming from the 
solar neutrino experiments.
We differentiate between two types of terrestrial 
experimental set-ups sensitive to vacuum oscillations. One 
which has its energy and baseline tuned to a maximum in the 
survival probability and another where the baseline ($L$) and energy 
($E$) would 
give a minimum in the survival probability.
We argue that sensitivity to $\theta_{12}$ increases 
substantially if the experiment is sensitive to a ``Survival Probability  
MINimum'' (SPMIN) instead of a ``Survival Probability MAXimum'' (SPMAX)
-- as is the case in KamLAND, and
propose a new reactor experiment which would give 
precise value of $\tan^2\thsol$ down to 14\%. 

We begin in Section 2 with a discussion of the potential of the 
experiments sensitive to different limits of the survival probability 
in constraining the mixing angle.
We then discuss the solar neutrino experiments
and delineate the impact of each one separately on the 
global allowed areas. We obtain bounds on \thsol from a future SNO NC data.
In the next section we introduce the KamLAND data and discuss how much the
uncertainty in $\theta_{12}$ is going to reduce with the
increased statistics in KamLAND.
We make a comparative study of various solar neutrino experiments along
with KamLAND data and determine the role of the individual experiments
in constraining $\theta_{12}$ and $\Delta m^2_{21}$.
The reasons for the low sensitivity of \kl to \thsol is expounded.
In Section 4 we propose a new reactor experiment which could 
in principle bring down the uncertainty in $\tan^2\thsol$ to 14\%.
In the next section we
examine the role of the future solar neutrino experiments -- Borexino 
and the LowNu experiments. We finally present our conclusions in 
Section 6.

\section{Solar Neutrino Experiments}

\begin{figure}[t]
\centerline{\epsfig{figure=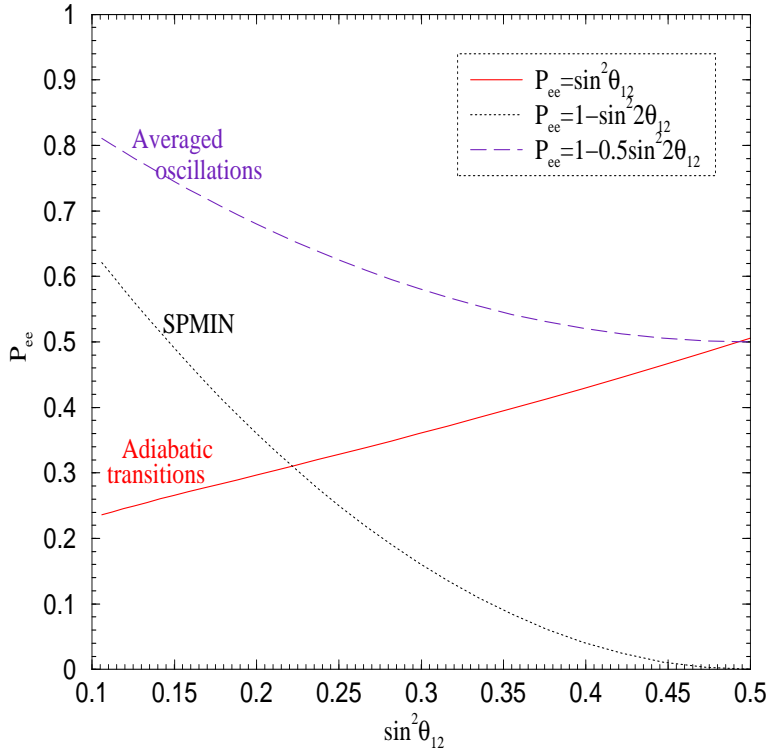,height=4.in,width=4.in}}
\caption{The survival probability $P_{ee}$ as a function $\sin^2\theta_{12}$ 
for $P_{ee}= \sin^2\theta_{12}$ (solid line), 
$P_{ee}= 1-0.5\sin^22\theta_{12}$ (dashed line) and 
$P_{ee}=1-\sin^22\theta_{12}$ (dotted line).}
\label{sens}
\end{figure}

The solar neutrinos come with a wide energy spectrum and 
have been observed on Earth in detectors with different 
energy thresholds. 
The survival probability for the low energy 
$pp$ (in Ga experiments -- SAGE, GALLEX and GNO) 
and the high energy \br fluxes (in SK and SNO)
in the now established 
LMA scenario can be very well approximated by
\be
P_{ee}({pp}) \approx 1-\frac{1}{2}\sin^22\thsol
\label{peepp}\\
P_{ee}({\br})\approx \sin^2\thsol+f_{reg}
\label{pee8b}
\ee 
where $f_{reg}$ is the $\nue$ regeneration inside the Earth. 
Thus the solar neutrinos in LMA are sensitive 
to \thsol, the degree of sensitivity depending on the energy of the 
relevant solar neutrinos observed. To expound this feature we 
present in Figure \ref{sens} the 
variation of $P_{ee}$ with $\sin^2\thsol$ for the different 
limits of the neutrino oscillation scenarios -- averaged oscillations 
(cf. Eq.(\ref{peepp})), 
fully adiabatic conversions in matter (cf. Eq(\ref{pee8b})) 
and ``full'' vacuum oscillations  
corresponding to 
``Survival Probability  
MINimum'' (SPMIN), that is 
$P_{ee}=1-\sin^22\thsol$\footnote{This  
case corresponds to vacuum oscillations with 
$\sin^2(\dm L/4E)\approx 1$ and we call this SPMIN, 
since $P_{ee}$ is minimum for this 
choice of $L/E$. 
The case where  $\sin^2(\dm L/4E)\approx 0$ is referred 
to in this paper as an SPMAX.}.
For both averaged oscillations and SPMIN 
the dependence of the probability 
is quadratic in $\sin^2\thsol$, while for complete adiabatic conversions (AD) 
the dependence is linear. Thus for the latter the error in 
$\sin^2\thsol$ is roughly same as the error in the probability $P_{ee}$. 
\be
(\Delta \sin^2\thsol)_{\rm AD} \sim \Delta P_{ee}
\label{ad}
\ee
While the corresponding error for averaged oscillations (AV) and 
SPMIN cases are roughly given by
\be
(\Delta \sin^2\thsol)_{\rm AV} \sim \frac{\Delta P_{ee}}{-2\cos 2\thsol}
\label{av}\\
(\Delta \sin^2\thsol)_{\rm SPMIN} \sim \frac{\Delta P_{ee}}{-4\cos 2\thsol}
\label{om}
\ee
the 
sensitivity to 
$\sin^2\thsol$ for averaged oscillations 
being reduced to roughly 1/2 of that for SPMIN.
We note from Eqs.(\ref{ad}), (\ref{av}) and (\ref{om}) that
for mixing angle not very close to maximal mixing, that is for 
$\cos2\thsol \gsim 0.25$ ($\sin^2\thsol \lsim 0.375$), 
the error in \thsol is least when we have a SPMIN. 
For $\cos2\thsol \gsim 0.5$ ($\sin^2\thsol \lsim 0.25$) 
even averaged oscillations are better 
suited for \thsol measurements than adiabatic conversions inside matter.
However for large mixing angles close to maximal, 
the adiabatic case has the maximum sensitivity.
All these features are evident in the Figure \ref{sens} which shows 
that  
for 
the SPMIN case
and for mixing not too close to maximal, the $P_{ee}$ has the 
sharpest dependence on the mixing angle and the \thsol sensitivity 
is maximum. Since the 99\% C.L. allowed values  
of \thsol is within the range $0.14 < \cos2\thsol < 0.57$, 
SPMIN seems most promising for constraining $\theta_{12}$.

\subsection{Bounds from current solar data}

While the Gallium (Ga) experiments, SAGE, 
GALLEX and GNO \cite{ga} are sensitive mostly to the $pp$ 
neutrinos, the SK \cite{Fukuda:2002pe}
and SNO \cite{Ahmad:2002jz} predominantly observe the higher energy \br
neutrino flux.
The Chlorine experiment (Cl) \cite{Cleveland:nv}
observes the intermediate energy 
\ber neutrinos in addition to the \br. 
Since the best-fit value for the mixing angle is large (with 
$\sin^2\theta_{12} \approx 0.3$), 
from the discussion above we expect SK and SNO 
to have a better handle over $\theta_{12}$. However the 
observed rates in the detectors depend not only on the survival 
probability but also on the initial solar neutrino flux in the Sun. 
The errors in the predicted fluxes are carried over to the errors 
in the parameters determined, reducing the net sensitivity.
While the $pp$ neutrinos are very accurately 
predicted and have theory errors of less than $\sim 1\%$, the \br neutrinos 
have a huge Standard Solar Model (SSM) 
uncertainty of $\sim 20\%$ \cite{bp00}. Thus on this  
front the ``sub-MeV'' experiments score over the higher energy 
solar neutrino experiments.

\begin{figure}[t]
\centerline{\psfig{figure=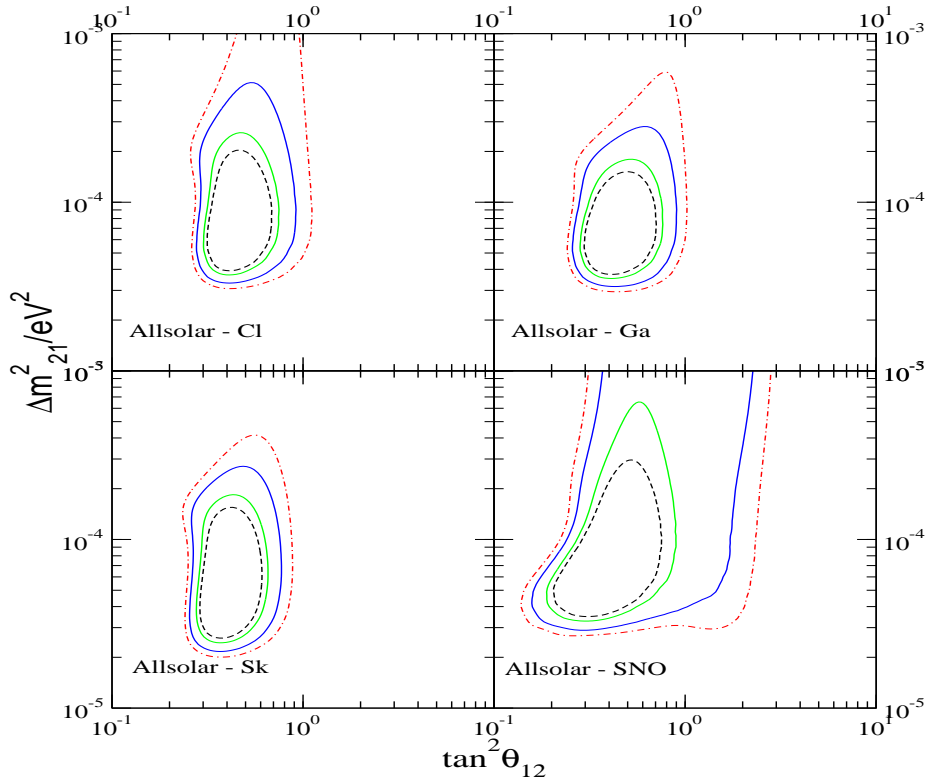,height=5.in,width=5.in}}
\caption{The 90\%, 95\%, 99\% and 99.73\% C.L. contours
from a $\chi^2$ analysis using all but one of the solar neutrino 
experiments. The experiment left out from analysis is indicated in the 
panels.
}
\label{allsol-1}
\end{figure}

 SK and SNO are real-time experiments and 
hence carry information regarding energy dependence of the suppression 
and potential matter effects as well.  
To project a realistic scenario of the potential of each of the 
solar neutrino experiments in constraining the parameters, we 
present in Figure \ref{allsol-1} the C.L. allowed 
contours \footnote{In our solar analysis we include the 
total rates from Cl and Ga, the full zenith angle spectral 
data from SK and the complete day-night spectral information from SNO
\cite{Bandyopadhyay:2002xj,Choubey:2002nc,Bandyopadhyay:2002qg}. 
Note that in the solar neutrino analysis the $^8B$ rates 
come as $f_B P_{ee}$ where $f_B$ is a normalization factor for the $^8B$ 
flux and is varied as a free parameter. }
from an analysis where all but 
one of the experiments is not considered\footnote{For the 
allowed regions from the individual solar neutrino 
experiment we refer to Figure 3 of 
\cite{Choubey:2002nc}.}. 
The figure shows that 
exclusion of Cl from the analysis raises the upper limit on both \dm and 
\thsol. 
Higher values of \dm and values of \thsol close to maximal mixing
give an energy independent suppression of the solar neutrino
flux within $\pm 10\%$ \cite{Choubey:2001bi}.
The Cl experiment with an observed rate that is $2\sigma$ away
from that predicted by maximal mixing disfavors these zones.
So omission of Cl makes these zones more allowed. 
SK is  consistent with
no energy dependence in the survival probability.
Thus SK
favors these quasi-energy independent
regions of the parameter space. 
The non-observation of any
significant day-night asymmetry in SK puts the lower bound on
$\Delta m^2_{21}$
and hence omission of SK loosens this bound.
The Ga observed rate of 0.55 is comparatively
closer to the rate predicted at maximal mixing ($=0.5$),
however the error in the $pp$ flux is only $\sim 1\%$ and this helps Ga to
disfavor maximal mixing.
Therefore excluding Ga slightly increases the upper limit of \thsol.  
But the strongest impact on the allowed regions of the parameter space comes
from SNO, which comprehensively rules out most of these
quasi-energy independent zones that predict a suppression rate
$P_{ee} \gsim 0.5$.
Thus without SNO the bounds become much weaker 
in both \dm and \thsol. The upper limit on \dm vanishes and the 
upper limit on \thsol becomes extremely poor, with large areas in the 
``dark zone'' (zones with $\thsol > \pi/4$) getting allowed.
Without SNO these areas were allowed since 
the 20\% uncertainty in the \br neutrino flux could be used to 
compensate for the higher survival probability and explain the 
global data. However with SNO the uncertainty in \br flux has come down to 
12\%, putting a sharp upper bound to both \dm ($\dm < 2.2\times 10^{-4}$ 
eV$^2$) and \thsol ($\tan^2\thsol <0.77 $) at 99\% C.L..

\subsection{Sensitivity of expected NC data from SNO}

\begin{figure}[t]
\centerline{\psfig{figure=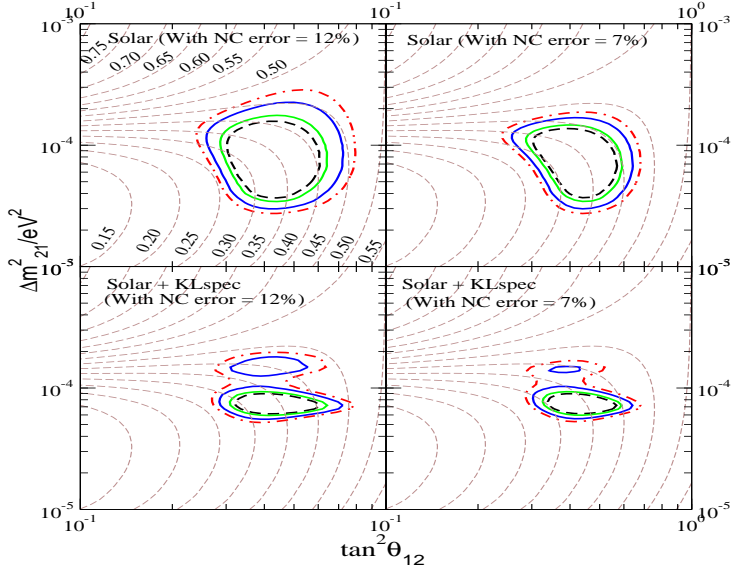,height=4.in,width=4.in}}
\caption{
The impact of the future SNO NC data (with error of 7\%) 
on the parameter space. The upper panels are for the global 
solar neutrino data with current (12\%) and future (7\%) error in NC.
The lower 2 panels are the corresponding allowed regions obtained 
by combining the \kl and the solar neutrino data.
The lines of constant CC/NC ratio in SNO are shown by dashed 
lines. 
}
\label{nc1}
\end{figure}

This tremendous power of SNO to  
constrain mass and mixing parameters
stems from its ability to 
simultaneously measure the neutrino suppression rate 
through the charged current (CC) interaction and the total 
\br neutrino flux 
through the independent neutral current (NC) measurement.
Thus by reducing errors in both $P_{ee}$ (from CC reaction) 
and the \br flux normalization $f_B$ (from the NC reaction), SNO 
can put better limits on the mass and mixing parameters. In particular 
it bounds the LMA zone in \dm from the top, chopping off parts of the 
parameter space for which the \br neutrinos do not undergo resonant 
transitions inside the Sun and therefore have a form of $P_{ee} 
\approx 1-0.5\sin^22\thsol$. These regions would give a $P_{ee}>0.5$ 
and could be accommodated with the CC data only if the initial \br 
flux was assumed to be less, or in other words $f_B<1$. However 
values of $f_B$ different from 1 are disfavored from the NC measurements 
of SNO and these high \dm regions get ruled out. Similarly in the 
adiabatic zone since $P_{ee}\approx \sin^2\thsol$, the larger values of 
$\sin^2\thsol$ close to maximal mixing would be 
allowed only if $f_B$ were to be assumed to be less than 1, which is 
at variance with the data as discussed above and hence these zones get 
severely constrained.

The upper left-hand panel of Figure \ref{nc1} shows the current C.L. allowed 
zones from the global solar neutrino experiments. Superimposed on them are 
the lines of constant CC/NC rates in SNO\footnote{Lines of constant 
day-night asymmetry in SNO are seen to be practically independent \thsol 
\cite{Maris:2001tg} and so we do not present them here. 
However they have a sharp \dm 
dependence which can be used as a consistency check on the \dm measurement 
from KamLAND.}. We note that the $3\sigma$ range of predicted CC/NC rates 
from the current solar limits are $0.23-0.47$. If SNO can measure a CC/NC 
ratio with smaller errors 
then the range for the allowed values
of \thsol would reduce.

The next phase of NC rate from SNO would come from capture of the 
neutron -- 
released in the neutral current breakup of heavy 
water -- on $^{35}Cl$ (salt). 
This data is expected to have much better 
statistics than the earlier data released last year, which was with neutron 
capture on deuterons. Since the efficiency of neutron captures on 
salt is about 83\% while that on deuterons only about 30\% we expect the 
statistical errors in the neutral current measurements to come down
to about 5\%.
It would be interesting to gauge how much the 
uncertainty in \thsol would reduce with better measurements of the 
total \br flux from SNO. Just to project the impact of reduced errors 
from SNO we show in the upper right-hand panel of Figure \ref{nc1} the 
allowed areas in the parameter space when the total error in the NC 
measurement is reduced to 7\%\footnote{The current systematic error 
in the NC data is about 9\%. However we make an optimistic reduction 
in the total errors in the future SNO NC measurements.}. Since the 
purpose of this figure is not accuracy but an 
optimistic projection of the impact of a futuristic 
SNO NC measurement, we have replaced the 34-binned SNO spectrum data used 
everywhere else in this paper, with the total charged and neutral current 
rates in SNO. The total rates are disentangled from the SNO spectrum data 
by assuming no spectral distortion for the \br flux.
Since we confine ourselves to the LMA zone where there is hardly 
any spectral distortion expected, we consider this to be an 
excellent approximation. We note that the limit on \thsol improves 
with reduced 
errors in NC and the 99\% C.L. bounds at $\dm=7\times 10^{-5}$ eV$^2$ reads 
$0.3 < \tan^2\thsol < 0.63$.

\section{KamLAND}

\subsection{Current Bounds}

\begin{figure}[t]
\centerline{\psfig{figure=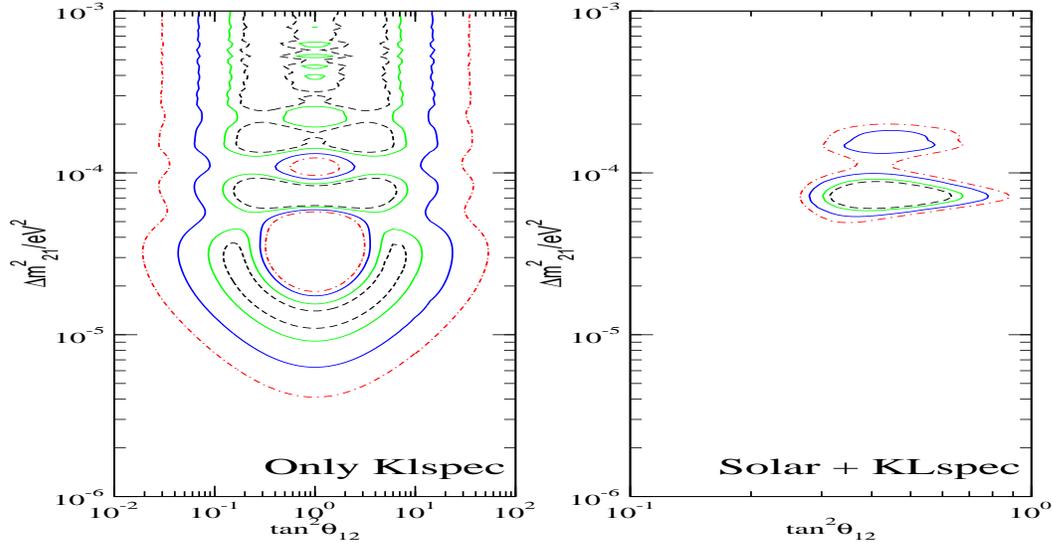,height=4.in,width=6.in}}
\caption{The 90\%, 95\%, 99\% and 99.73\% C.L. contours
from a $\chi^2$ analysis using
the KamLAND spectrum data alone (left panel) and 
the combined \kl and global solar data (right panel).
}
\label{klsol}
\end{figure}

\begin{figure}[t]
\centerline{\psfig{figure=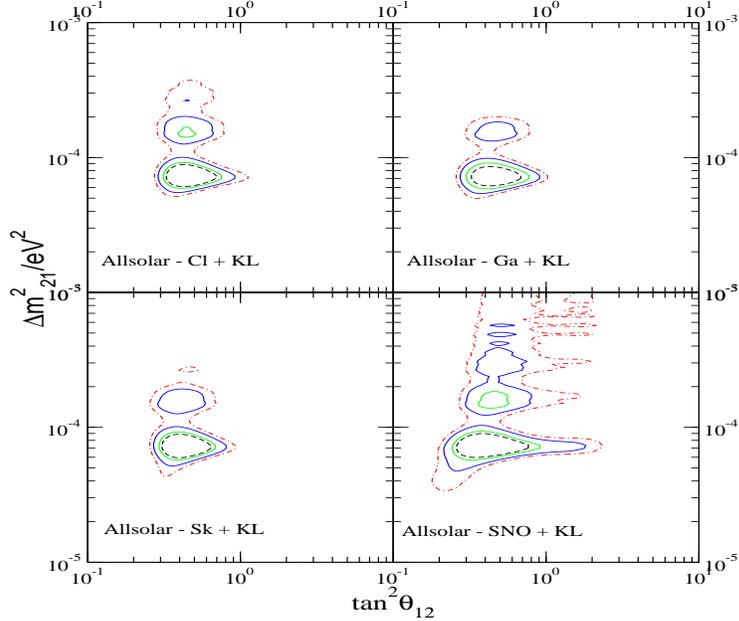,height=4.in,width=4.in}}
\caption{
Same as Figure \ref{allsol-1} but with the \kl data included.} 

\label{all-1}
\end{figure}


%
%
\begin{figure}[t]
\centerline{\psfig{figure=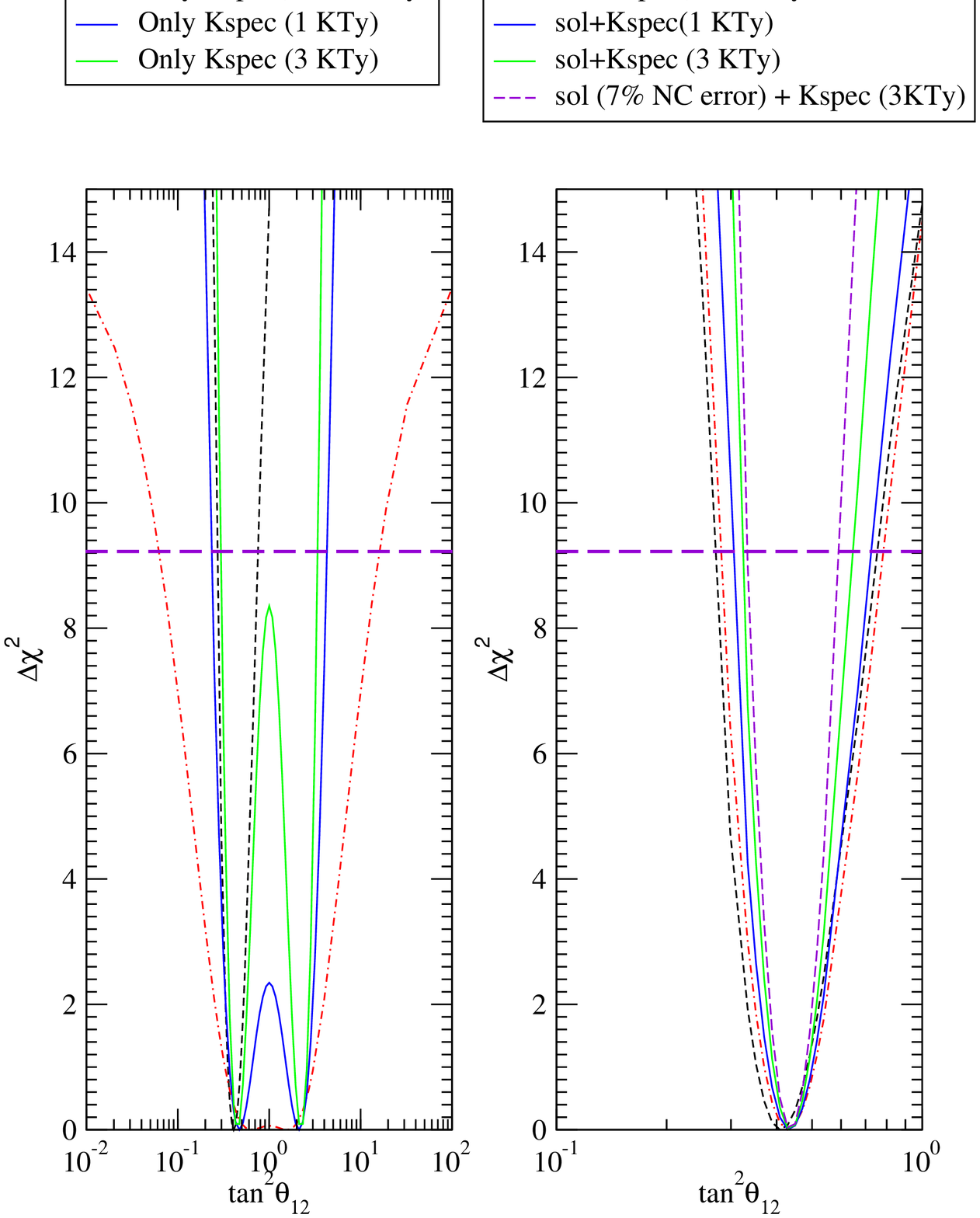,height=5in,width=6in}}
\caption{$\Delta \chi^2$ versus $\tan^2\thsol$ for only \kl data 
(left panel) and combined \kl and Solar neutrino data (right panel).
For \kl we use the declared 0.162 kTy data as well as the 1 kTy 
and 3 kTy projected spectral data, with the spectrum simulated at 
the low-LMA best-fit point. The long-dashed line gives the 99\% C.L. 
bound for 2 parameter fit.
}
\label{delchi}
\end{figure}

After the announcement of the first KamLAND results 
\cite{Eguchi:2002dm} there was a plethora 
of papers discussing the impact of \kl on the mass and mixing parameters, 
$\Delta m_{\odot}^2$ and $\theta_\odot$ 
\cite{Bandyopadhyay:2002en,others}.
The \kl spectrum even though still short on statistics, is 
powerful enough to disintegrate the solar neutrino parameter space 
into two disjoint islands at the 99\% C.L. -- one around the global 
best-fit of $\dm=7.17\times 10^{-5}$ eV$^2$ and $\tan^2\theta_{12}=0.44$ 
and another 
at $\dm=1.49\times 10^{-4}$ eV$^2$ 
and $\tan^2\theta_{12}=0.43$. We call them low-LMA 
and high-LMA respectively. High-LMA appears at a reduced statistical 
significance of about $2\sigma$. The two islands however join at the 
$3\sigma$ level. We show the currently allowed zones in 
Figure \ref{klsol}\footnote{For the \kl analysis 
we take the 13-binned spectrum data. We assume 
Poisson distribution for the \kl spectrum. For the solar neutrino   
data the error analysis assumes a Gaussian distribution.
For the details of our solar neutrino 
and \kl analysis techniques and codes we refer the reader to 
\cite{Bandyopadhyay:2002xj,Choubey:2002nc,Bandyopadhyay:2002en,
Bandyopadhyay:2002qg}.}. The right-hand panel of this figure gives the 
allowed areas from the \kl data alone, while the left-hand panel gives 
the combined allowed zones from solar and \kl data. 
From a global analysis involving the solar and \kl data 
the $3\sigma$ ranges are
\be
0.27&<~\tan^2\theta_{12}~<&0.88\\
4.96\times 10^{-5}&<~ \Delta m_{21}^2~ <& 2.0\times10^{-4}
\ee
The 99\% range for the parameters are \cite{Bandyopadhyay:2002en},
\be
0.28&<~\tan^2\theta_{12}~<&0.79\\
5.3\times 10^{-5}&< ~\Delta m_{21}^2 ~<& 9.9\times10^{-5},~~~({\rm low-LMA})
\ee
\be
0.34&<~\tan^2\theta_{12}~<&0.55\\
1.3\times 10^{-4}&<~ \Delta m_{21}^2 ~<& 1.8\times10^{-4},~~~({\rm high-LMA})
\ee
We note that low-LMA allows a much larger range of \thsol than high-LMA.
This has more to do with the fact that the global best-fit is in 
low-LMA than anything else. If the contour at high-LMA was to be plotted 
with respect to the local minima at high-LMA, then the allowed range 
of \thsol would be almost the same.

To study the impact of each of the solar neutrino experiments 
in determining the allowed range of the mixing parameters 
in conjunction with KamLAND, 
we show in 
Figure \ref{all-1} the allowed areas from a 
combined analysis involving the \kl data and the solar data, with
each panel showing the areas obtained when one of the solar neutrino 
experiments is {\it excluded}. 
The Figure shows that 
neglecting Cl helps to make the high-LMA slightly more allowed
and the 3$\sigma$ contour extends to larger $\Delta m^2$ while omission of
Ga and SK does not change the contours much with respect to the 
global contours of figure. 4. However the exclusion 
of SNO completely removes the upper bound on \dm and allows \thsol 
to move into the ``dark zone'' even at 99\% C.L.. 
This again exemplifies the power of SNO in 
constraining the quasi-energy independent zones as discussed in the 
previous section. In fact we have checked that 
SNO alone combined with KamLAND, can almost restrict both \dm and \thsol 
within the current global allowed range.

\subsection{Reduced SNO NC errors and \kl}

It would be interesting to check if the \kl data with future 
solar neutrino data in general and SNO NC data in particular, could 
improve the limits on the parameters or not.
The lower 2 panels in Figure \ref{nc1} 
show the impact of the next phase 
SNO NC data in conjunction with the \kl data. 
The lower left panel of 
the figure shows the 
current global allowed regions obtained from the combined solar 
and \kl data. Also shown superimposed are the 
constant lines for the CC/NC rates in SNO.
The predicted $3\sigma$ range for the CC/NC rates is 
seen to be $0.27 - 0.47$.
The lower right hand panel gives the allowed areas obtained when the error 
in NC measurement is reduced from 12\% to 7\% as discussed earlier.
We again reiterate that for this figure with future SNO NC measurement 
we have used the CC and NC rates instead of the full SNO day-night spectrum 
used in the rest of the paper.
The combination of the solar with reduced NC errors and \kl is seen 
to constrain \thsol to $0.3 < \tan^2\thsol < 0.63$ at 99\% C.L., 
which is the same as that obtained without \kl and with improved NC.
Thus we again 
note that inclusion of the current \kl data in the global analysis 
brings no improvement on the limits for \thsol. 
The results obtained from a combined future SNO NC and future \kl data are 
presented in the following sections.

\subsection{Sensitivity of projected \kl data}

In \cite {Bandyopadhyay:2002en} we made a projected 
analysis using the 1 kTy \kl spectrum simulated at some strategic 
points in and around the high-LMA and low-LMA allowed regions
and probed the potential of a 
statistics enriched \kl data sample to plump for the right solution
between the two. The 3 kTy \kl data is obviously expected 
to further tighten the bounds on the mixing parameters 
\cite{Bandyopadhyay:2002mc}. 
The sensitivity of \kl to \dm is found to be remarkable. 
To study the limits that \kl would be expected to 
impose on the mixing angle \thsol 
with more statistics, we 
present in 
Figure \ref{delchi} the $\Delta \chi^2 (=\chi^2-\chi^2_{min})$ 
as a function of $\tan^2\thsol$, keeping \dm free. The left-hand panel gives 
the limits obtained from \kl data alone, with the declared 0.162 kTy data and 
the projected 1 kTy and 3 kTy data, 
simulated at the current low-LMA best-fit point.
The right-hand panel gives the corresponding bounds when \kl is combined with 
the solar data. 
The limits on the value of \thsol will 
depend somewhat on the point in the parameter space where the 
projected \kl spectra are simulated. We present here just the bounds 
obtained if the future \kl spectrum sticks to its current trend and 
roots for the low-LMA best-fit point.
Also shown in both the panels is the curve corresponding to the 
global solar neutrino data alone. 

Apart from the increased statistics we have also studied the
role of the reduced systematics on the allowed parameter ranges.
The current 0.162 kTy \kl data has a rather large 
and very conservative systematic error of 6.42\% \cite{Eguchi:2002dm}.
However the \kl collaboration hopes to improve their systematics in the future.
The bulk of the systematic error  comes from
the error in the knowledge of the fiducial volume which could
be improved by making calibration measurements.
For 1 kTy data the systematic error could reduce to the 5\% level
with better understanding of the detector and more statistics.
For the 3 kTy data sample the systematic uncertainties could be lowered to
even 3\% with three-dimensional calibrations and better understanding 
of reactor neutrino 
flux\footnote{We thank Professor Atsuto Suzuki, Professor 
Fumihiko Suekane and Professor Sandip Pakvasa for 
information regarding the most optimistic 
estimates on the possible future systematic errors in KamLAND.}. 
We have assumed an expected   
5\% systematic uncertainty for our analysis with the
1 kTy \kl data and a more optimistic 
3\% systematic uncertainty for the 3 kTy data 
sample.

From the Figure \ref{delchi} we see that 
even with 3 kTy statistics (and with only 3\% systematic error), \kl 
would   
constrain \thsol only marginally better than the 
current solar neutrino 
experiments. 
Also, \kl being insensitive to matter effects 
has a $\thsol$ and $\pi/2 - \thsol$ ambiguity and therefore allows 
regions on both side of maximal mixing. 
Maximal mixing itself cannot be ruled out 
by the 3 kTy \kl data alone.
The right-hand panel shows that the combined \thsol limits from 
the global solar neutrino 
data and future \kl data, would be somewhat
more constricted than that obtained from the current 
solar data alone. Also shown in the right-hand panel of Figure 
\ref{delchi} is the \thsol sensitivity curve obtained by combining 
the 3 kTy \kl data (with 3\% systematic uncertainty) with the global 
solar neutrino data, where the total uncertainty in the SNO NC data 
has been reduced from 12\% as of now, to only 7\% expected from a future 
SNO measurement. Reduction of the SNO NC error reduces the 
combined allowed \thsol 
range as discussed in Section 3.2, particularly on the large mixing side.

\begin{table}
\begin{center}
\begin{tabular}{ccccccccc}
\hline\hline
Data & 99\% CL &99\% CL  &1 $\sigma$ & 2$\sigma$ & 99\% CL 
& 1 $\sigma$ & 2 $\sigma$ & 99\% CL \cr
set & range of & spread & range  &range  &range  & spread  
& spread & spread  \cr
used & $\dm\times$ & of & of &of & 
of & in  
& in & in \cr
 & 10$^{-5}$eV$^2$ & \dm & $\tan^2\theta_{12}$ 
& $\tan^2\theta_{12}$ & $\tan^2\theta_{12}$ & $\tan^2\theta_{12}$
& $\tan^2\theta_{12}$ & $\tan^2\theta_{12}$ \cr
\hline\hline
only sol & 3.2 - 24.0  
& 76\%& $.33-.53$ & $.29-.66$ & $.27-.75$ & 23\% & 39\% & 47\% \cr
sol+162 Ty &  5.3 - 9.9 
& 30\% & $.34-.55$ &$.30-.68$ & 
 $.28 -.78$ & 23\% & 39\%  & 47\%  \cr
sol+1 kTy & 6.7 - 8.0  
& 9\% & $.36-.54$
& $.33-.65$ & $.30-.72$ & 20\% & 33\%  & 41\% \cr
sol+3 kTy & 6.8 - 7.7 
& 6\% & $.37-.52$ & $.34-.59$ & $.33-.65$ & 17\% & 27\% & 33\% \cr
sol(7\%)+3 kTy & 6.8 - 7.7
& 6\% & $.38-.50$
& $.35-.56$ & $.33-.60$ & 14\% & 23\%  & 29\% \cr

\hline\hline
\end{tabular}
\label{klbounds}
\caption
{The range of parameter values allowed and the corresponding spread. 
For the current 
observed solar+\kl analysis we show the ranges and the spread only in the 
low-LMA region. For the 1 kTy and 3 kTy ranges we have simulated the 
spectrum at the current low-LMA best-fit. We assume 5\% systematic error
for 1 kTy \kl spectrum and 3\% systematic error for 3 kTy \kl spectrum. The 
last row of the Table corresponds to a combination of the 3 kTy \kl data 
and the global solar neutrino data where the SNO NC error has been reduced 
to only 7\%.
}
\end{center}
\end{table}

In Table 1 we explicitly present the 99\% C.L. allowed ranges 
for the solar neutrino parameters in low-LMA,
allowed from combined solar and KamLAND\footnote{For 
the various  C.L. limits in the Table 1 we take $\Delta\chi^2$ 
corresponding 
to a two parameter fit.}.
Shown are the current bounds on \dm and $\tan^2\thsol$ and those 
expected after 1 kTy and 
3 kTy of \kl data taking. The sensitivity of \kl to \dm is tremendous.
Since the thrust of this paper is to study the limits on the solar 
mixing angle,
we also give the $1\sigma$ and $2\sigma$ limits for $\tan^2\thsol$.  
Also shown are the \% spread in the oscillation 
parameters. We define the ``spread'' as
\be
{\rm spread} = \frac{ prm_{max} - prm_{min}}
{prm_{max} + prm_{min}}\times 100
\label{error}
\ee
where ${prm}$ denotes the parameter \dm or $\tan^2\thsol$.
\kl is extremely good in pinning down the 
value of \dm. The ``spread'' in \dm comes down from 30\% as of now 
to 9\%(6\%) with 1 kTy(3 kTy) \kl spectrum data. However 
its sensitivity to \thsol is not of the same order. The spread in 
$\tan^2\thsol$ goes down only to 41\%(33\%) from 47\% with the 
1 kTy(3 kTy) \kl spectrum data combined with the solar data. 
Thus as discussed before, even with the most optimistic estimates for the 
\kl error analysis, the sensitivity of \kl to \thsol 
is not much better than of the current solar data and the 
range of allowed value for \thsol does not reduce by a large amount 
even after incorporating KamLAND.

The last row of Table 1 shows the allowed range of parameter 
values from a combined analysis of the 3 kTy \kl data (with 3\% 
systematic uncertainty) and the 
global solar data, where the total error in the SNO NC data has 
been reduced to 7\%.
We note that this combination of futuristic as well as optimistic 
expected data from SNO NC and \kl reduces the \thsol uncertainty 
to 29\% at the 99\% C.L.. However if we compare the range of allowed 
values for \thsol given in the last row of Table 1 with that 
obtained from an analysis of only the solar data with 
SNO NC error of 7\% given in the previous section 3.2, we note 
that solar data alone with improved SNO NC measurements can reduce the 
spread in \thsol to 35\% at 99\% C.L.. Thus even in this scenario
inclusion of the \kl data 
helps in reducing the $\tan^2\thsol$ spread only from 35\% to 29\%, and 
even 29\% is large when compared with the 6\% spread expected for 
\dm from \kl alone.

The reactor antineutrinos do not have any 
significant matter effects in \kl and hence the survival probability has 
the vacuum oscillation form 
\be
P_{ee}=1-\sum_i \sin^22\thsol \sin^2\left(\frac{\dm L_i}{4E}\right)
\label{probkl}
\ee
where $L_i$ stands for the different reactor distances. As discussed in 
Section 2, experiments sensitive to averaged vacuum oscillation 
probability are less sensitive to \thsol, particularly close to 
maximal mixing. However in \kl the probability, even though partially averaged 
due summing over the various reactor distances, is not completely averaged. 
The \kl spectrum shows a peak around 3.6 MeV which is well reproduced 
by $\dm \sim 7.2 \times 10^{-5}$ eV$^2$. This sensitivity to shape 
gives \kl the ability to accurately pin down \dm.

However the sensitivity of \kl to \thsol around the best-fit point
is actually worse than experiments which observe only 
averaged oscillations. The reason being that the \kl 
data is consistent with a ``survival probability maximum'' (SPMAX) 
of vacuum oscillations, 
with an oscillation peak in the part 
of the neutrino spectrum that is statistically most relevant. 
At SPMAX the \dm dependent 
$\sin^2(\dm L_i/4E)$ term is close to zero, 
smothering any \thsol dependence along with it. 
As discussed in Section 2,
the \thsol sensitivity 
would have been more, had the \kl distances been 
tuned to a SPMIN.

\section{A new reactor experiment for $\theta_\odot$?}
 
\begin{figure}[p]
\centerline{\psfig{figure=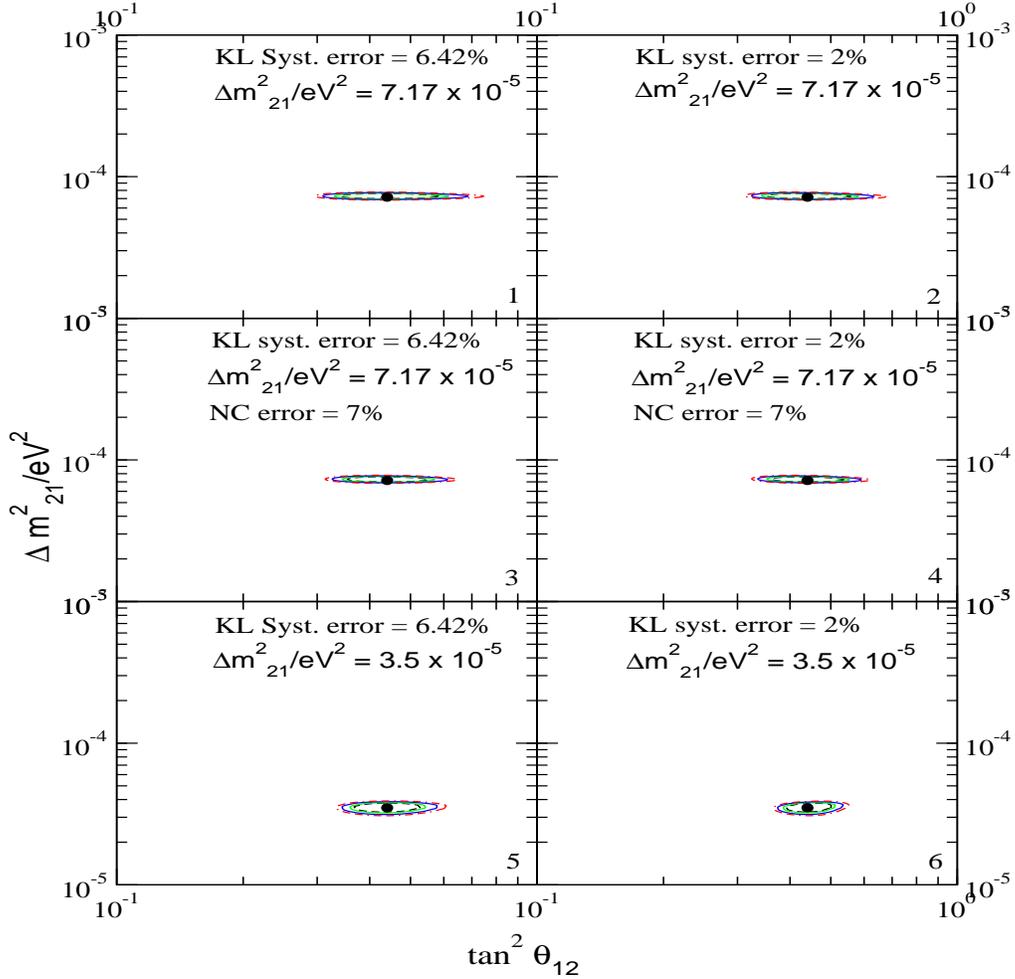,height=6in,width=5.5in}}
\caption{The 90\%, 95\%, 99\% and 99.73\% C.L. contours for
the combined analysis using the solar and 3 kTy
projected KamLAND spectrum.
The upper panels are for the simulated \kl 
spectrum at low-LMA best-fit 
parameters and the current solar data, the middle panels are for 
the simulated \kl 
spectrum at low-LMA best-fit parameters and the solar data with 
future SNO NC data (7\% error),
while the lower panels are for \kl data simulated at 
$\Delta m_{21}^2=3.5\times 10^{-5}$ 
eV$^2$ and $\tan^2\theta_{12}=0.44$ and the current global solar data. 
The left-hand panels use the current  
\kl systematic uncertainty of 6.42\% while the right-hand panels 
correspond to a fictitious systematic uncertainty of just 2\%.
}
\label{fict}
\end{figure}

From the Figure \ref{sens} presented in Section 2 and the 
discussion on \kl sensitivity to \thsol in the previous section 
we conclude that a reactor experiment can measure \thsol accurately 
enough only if it is sensitive to the SPMIN.
To further elaborate our point 
in figure \ref{fict} we present the allowed areas  at 7.2$\times 10^{-5}$ 
eV$^2$ (SPMAX) 
and for a 
fictitious spectrum data in \kl 
simulated at $\dm = 3.5\times 10^{-5}$ eV$^2$ --  
which corresponds to an effective SPMIN
in \kl.  
We show limits for the current \kl
systematic uncertainty of 6.42\% and a systematic 
uncertainty of just 2\% {\footnote { We want to emphasise that the 
2\% uncertainty consdiered in figure \ref{fict} is a fictitious value 
 -- the 3\% systematic error in \kl after 
3 kTy of data which we assume in the previous section is already the
most optimistic estimate. 
However we present the contours for this fictitious case in order to 
facilitate comparison with scenarios presented later in this section.}}. 
We take 3  statistics for \kl in all the cases. 
The \% spread in uncertainty for the SPMAX case with 6.42\% 
systematic uncertainty is  37\% while for the SPMIN case with 
the same systematics the spread is 25\%. 
The effect of reducing
the systematics to 2\%  results in the spread coming down to 
32\% and 19\% respectively. 
We have also explored the effect of reducing the SNO NC 
error to 7\%  for  the SPMAX case. 
The resulting contours are presented in the middle panels 
of figure \ref{fict}. 
The $\tan^2\thsol$ spread for this case 
with  2\% systematic error in \kl  
is 27\%. This emboldens us to believe that the 
most suitable experiment for \thsol measurement is an experiment 
sensitive to the SPMIN as expected in Section 2. 

Thus unprecedented sensitivity to \thsol can be achieved in a 
terrestrial experiment if the distance traveled by the neutrino 
beam is tuned so that the detector observes a complete vacuum 
oscillation. 
The oscillation wavelength of the neutrinos can be calculated 
with reasonable accuracy with information on \dm from \kl.
For a reactor experiment which has a large flux around $3-4$ MeV, 
the detector needs to be placed at about 70 km from a powerful 
nuclear reactor in order to be sensitive to the oscillation 
SPMIN\footnote{Here we assume that the current best-fit \dm in low-LMA 
is the right value.}. Also important for accurate \thsol determination is 
to reduce the systematics. The major part of the 6.42\% error in \kl 
comes from sources which affect the overall normalization of the 
observed anti-neutrino spectrum. These can be reduced if the 
experiment uses the near-far detector technique in which there are 
two identical detectors, one close to the reactor and 
another further away \cite{Kozlov:2001jv,Minakata:2002jv}. 
Comparison of the number of detected events in the 
two detectors can be then used to reduce the systematics.

\begin{figure}[t]
\centerline{\psfig{figure=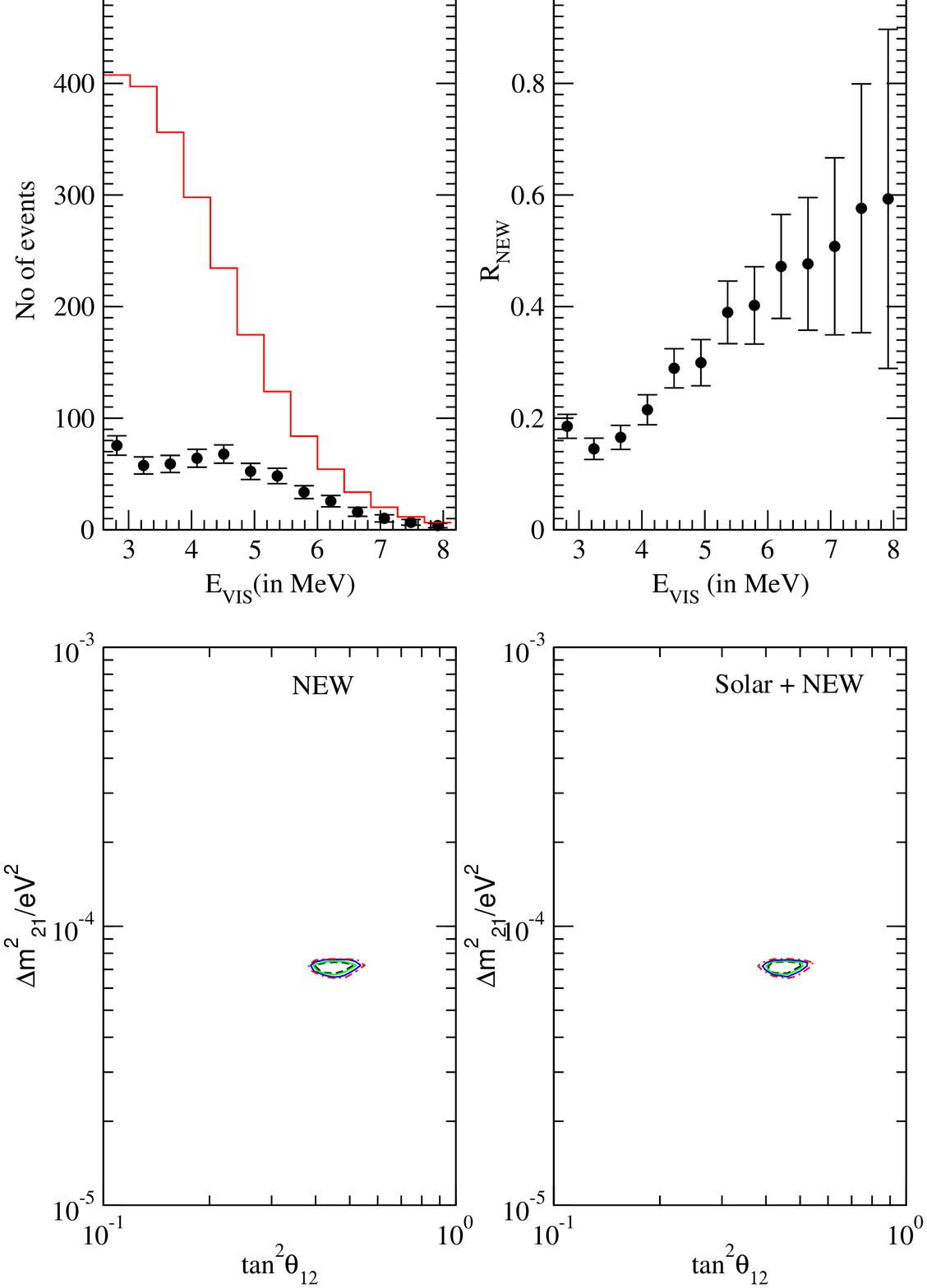,height=4.in,width=5.in}}
\caption{The simulated 3 kTy spectrum data at the low-LMA 
best-fit point and the allowed areas in the $\dm-\tan^2\thsol$ 
parameter space
for a 24.6 GWatt reactor experiment with 
a baseline of 70 km. The top-left panel gives the 
simulated spectrum data and the expected events, shown by the histograms.
The top-right panel shows $R_{NEW}$, the corresponding ratio  
of the ``data'' to expected 
events as a function of the visible energy. The bottom-left panel 
gives the allowed areas obtained using just the new reactor experiment.
The bottom-right panel presents the allowed areas from the 
combined solar and new reactor experiment data.
}
\label{fictnew}
\end{figure}

We show in Figure \ref{fictnew} the constraints on the mass and 
mixing parameters obtained using a ``new'' reactor experiment 
whose baseline is tuned to an oscillation SPMIN.
We use the antineutrino flux from a reactor a la 
Kashiwazaki nuclear reactor in Japan with a maximum
power generation of about 24.6 GWatt. We assume a 80\% 
efficiency for the reactor output and simulate the 3 kTy data 
at the low-LMA best-fit for a \kl like detector placed at 70 km from 
the reactor source and which has systematic errors of only 2\%. 
The top-left panel of the Figure \ref{fictnew}
shows the simulated spectrum data. The histogram shows the expected 
spectrum for no oscillations. 
$E_{vis}$ is the ``visible'' energy of the scattered electrons.
The top-right panel gives 
the ratio of the simulated oscillations to the no oscillation numbers.
The sharp minima around $3-4$ MeV is clearly visible.
The bottom-left panel gives the C.L. allowed areas obtained from 
this new reactor experiment data alone. With 3 kTy statistics we find a 
marked improvement in the \thsol bound with the 99\% range 
$0.39 < \tan^2\thsol < 0.52$ giving a spread of 14\%.\footnote{Note that the
first panel on the bottom line of Figure \ref{fictnew} admits a  mirror 
solution on the ``dark side'' because of the $\thsol-(\pi/2-\thsol)$ 
ambiguity in all experiments sensitive to oscillations in vacuum. 
This dark side solution can be ruled out by including the 
solar neutrino data.}

\section{Other future experiments}

We briefly discuss the sensitivity of the some of other 
next generation solar neutrino experiments. The most 
important among them are the Borexino which is sensitive 
to the monochromatic \ber neutrinos coming from the Sun and 
the sub-MeV solar neutrino experiments -- the so called 
LowNu experiments.

\subsection{Borexino}

Borexino is a 300 ton organic liquid scintillator detector, 
viewed by 2200 photomultiplier tubes \cite{borex}. 
The Borexino detector due to start operations soon,  
has achieved a background reduction at sub-MeV energies never 
attempted before in a real time experiment.
Borexino is tuned to detect mainly the \ber solar neutrinos by the 
elastic $\nu-e$ scattering process. The detector will operate in the 
electron recoil energy window of $0.25-0.8$ MeV to observe the 
mono-energetic 0.862 MeV \ber line which 
scatter electrons with a Compton edge at 0.66 MeV, the edge being somewhat 
smeared by the energy resolution of the detector. 

\begin{table}
\begin{center}
\begin{tabular}{c|ccc|c}
\hline\hline
Solution & $R_{Be}^{BF}$&$R_{Be}^{max}$&$R_{Be}^{min}$ & $A_{DN}$\cr
\hline
low-LMA & 0.65 & 0.71 & 0.61 & 0.04\cr
high-LMA & 0.66 & 0.71 & 0.63 & 0.01\cr
\hline\hline
\end{tabular}
\label{be2}
\caption
{The best-fit and $3\sigma$ range of predicted values for \bx
for the low-LMA and high-LMA solutions. Also shown is the value of 
the day-night asymmetry expected.}
\end{center}
\end{table}
\begin{figure}[t]
\centerline{\epsfig{figure=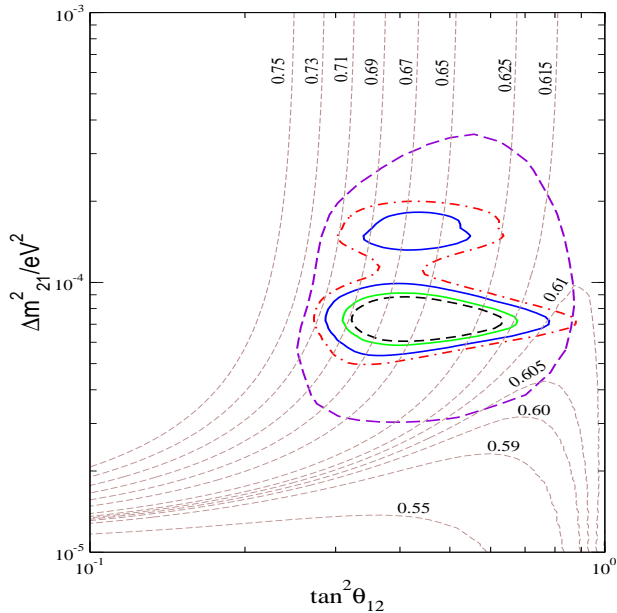,height=4.in,width=3.5in}}
\caption{The isorate lines for the \bx detector in the
$\Delta m_{21}^2-\tan^2\theta_{12}$ plane.
Also shown are the C.L. contours from the global analysis 
of the solar and the \kl data. 
Also shown by the purple dashed line 
is the only solar $3\sigma$ contour.
}
\label{beiso}
\end{figure}

We present in Figure \ref{beiso} the 
lines of constant \bx rate in the $\Delta m_{21}^2-\tan^2\theta_{12}$ LMA zone.
The \bx rate $R_{Be}$ is defined as the ratio of the value predicted 
by oscillations to the no oscillation SSM value.  
The global allowed 90\%, 95\%, 99\% and 
99.73\% C.L. contours are  
shown. Superimposed is the $3\sigma$ contour from the analysis of the 
only solar data. In Table 2 we show the 
predicted rate in \bx for the low-LMA and high-LMA best-fit solutions and 
the corresponding $3\sigma$ ranges. From Figure \ref{beiso} and Table 
2 we note that \bx in the LMA zone has almost no sensitivity 
to $\Delta m_{21}^2$. The reason being that for very low values of neutrino 
energies the solar matter effects are negligible while for \dm in the 
LMA zone there are hardly any Earth matter effect. Hence 
the survival probability can be approximated by averaged 
oscillations (cf. Eq.(\ref{peepp})).
Therefore \bx is not expected to sharpen our knowledge of \dm any further. 
Even the \thsol dependence is rather weak. 
This is due to the fact that the survival probability is 
of the averaged vacuum oscillation form which as discussed in Section 2
reduces the sensitivity of \bx to \thsol.

The $3\sigma$ error in the predicted value of \bx rate given in 
Table 2 from the current 
information on the parameter ranges is $\pm 0.06$. 
The corresponding 
$1\sigma$ range is $0.63< R_{Be} < 0.68$ implying an uncertainty of 
about $\pm 0.02$. 
Since there is 
hardly any \dm dependence involved the entire range can be attributed 
to the current uncertainty in \thsol. \bx could improve on the 
\thsol uncertainty if it could measure $R_{Be}$ with a $1\sigma$ error 
less than about 0.02. The low-LMA predicts about 13,000 events in 
\bx after one year of data taking. This gives a statistical error of about 
0.9\% only. However \bx {\it may} still 
have large errors coming from its background 
selection. 

In Table 2 we have also shown the day-night asymmetry 
expected in \bx for the currently allowed parameter values.
\bx will see no difference between the event rates at day and 
during night. Until the recent results from \kl the major role 
which \bx was expected to play was to give ``smoking gun'' signal 
for the low $\Delta m_{21}^2$ solution
LOW by observing a large day-night asymmetry and for 
the vacuum oscillation solution 
by observing seasonal variation of the \ber flux. The large 
day-night asymmetry expected due to the small energy sensitivity 
of \bx and the immaculate control over seasonal effects coming from 
the fact that \ber is a mono-energetic line -- not to mention 
its ability to pin down the SMA solution which predicted almost 
no $\nue$ events in \bx.
However all three  
are comprehensively disfavored now. 
Unfortunately the only region of parameter space where \bx 
lacks strength is the LMA, which is the correct solution to the 
solar neutrino problem.

\subsection{Low-Nu experiments}

\begin{figure}[t]
\centerline{\epsfig{figure=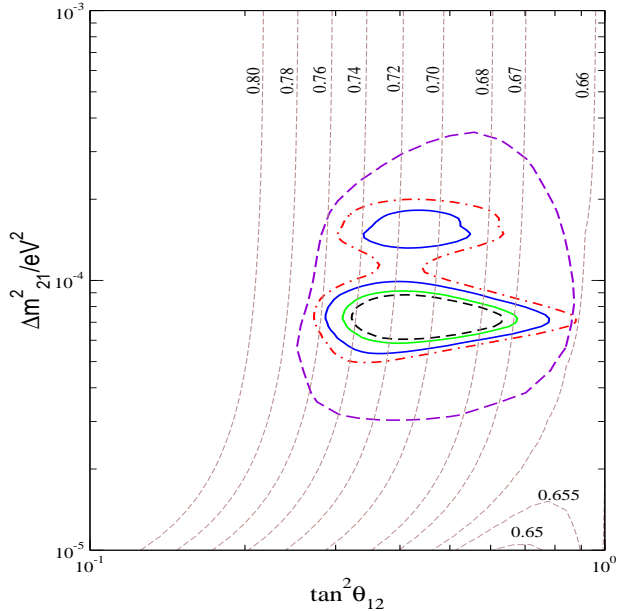,height=4.in,width=3.5in}}
\caption{The isorate lines for a generic $pp$ - e scattering 
experiment in the
$\Delta m_{21}^2-\tan^2\theta_{12}$ plane.
Also shown are the C.L. contours from the global analysis 
of the solar and the \kl data. 
Also shown by the purple dashed line 
is the only solar $3\sigma$ contour.
}
\label{lownuiso}
\end{figure}

There are a number of planned sub-MeV solar neutrino experiments 
which will look to observe the $pp$ flux using either charged 
current reactions (LENS, MOON, SIREN \cite{lownu}) or 
electron scattering process(XMASS, CLEAN, HERON, MUNU, GENIUS 
\cite{lownu}) for detecting the $pp$ neutrinos. 
While each of these electron scattering 
experiments use different detection techniques, the basic 
reaction involved is the scattering of the $pp$ neutrinos 
off the electrons in the detector. Hence we present in 
Figure \ref{lownuiso} the lines of constant rate predicted in a 
{\it generic} LowNu electron scattering experiment. Again we 
note that the $iso-pp$ rates have very little \dm dependence. 
The $3\sigma$ 
range predicted for $pp-e$ scattering is $0.66 - 0.76$. 
The corresponding  $1\sigma$ predicted range is $0.68 -0.73$.
The advantage
that these experiments have is that the $pp$ flux is 
predicted to within 1\% accuracy.
Thus the LowNu experiments may 
have a fair chance to pin down the value of the mixing 
angle $\theta_{12}$, {\it if} they can keep down their experimental errors.

\section{Conclusions}

With both solar and atmospheric neutrino oscillations confirmed the 
next turn in the research in neutrino physics is towards the precision
determination of the oscillation parameters of the PMNS matrix. 
In this paper we explore in detail how accurately the current and future 
experiments will be able to predict $\theta_{\odot}$ ($\theta_{12}$) and 
show that with the current set of experiments  
the uncertainty level in the determination of $\theta_{12}$ 
may stay well 
above the desired 10\% level (at 99\% C.L.).  
The spectrum data from the  
KamLAND experiment with only 0.162 kTy exposure in conjunction with the 
global solar data reveals an 
unprecedented sensitivity in constraining $\Delta m^2_{21}$, reducing 
the 99\% C.L. spread in  
$\Delta m^2_{21}$ to 30\% as compared to 76\% 
allowed by global solar data.   
A projected 
analysis with 3 kTy of simulated spectrum at the present best-fit 
reveals that the uncertainty 
in $\Delta m^2_{21}$ can be brought down to the $<$ 10\% level. 
However even with 3 kTy of exposure the $\tan^2\theta_{12}$ can hover in a 
$\sim 33\%$ uncertainty range.  
 
We make a comparative study of the $\theta_{12}$ sensitivity of the various 
solar neutrino experiments and KamLAND.
The sensitivity of an experiment 
to \thsol depends on the form of the survival probability relevant 
for it.  Thus the 
$\theta_{12}$ sensitivity of the solar neutrino experiments
are linked with the 
neutrino energy 
threshold. In SK and SNO, the high energy neutrinos are observed and the 
solar neutrinos undergo adiabatic transformation
($P_{ee} \sim f_B \sin^2\theta_{12}$)
resulting in an increased 
theta sensitivity as compared to the experiments which are sensitive to 
low energy neutrinos for which the survival probability is of the form 
$P_{ee} = 1 - 0.5 \sin^2 2\theta_{12}$. 
SNO has a better control over $\theta_{12}$ than SK as it is sensitive to the 
total $^8B$ flux through its neutral current channel and hence limits the 
range of $f_B$, the $^8B$ flux normalization to 12\%.
We make a projected sensitivity test for the future SNO NC measurement and 
get the limits on $\theta_{12}$. 

For the low energy neutrinos detected by the KamLAND  
detector the matter effects are absent. 
Therefore the relevant probability is the vacuum oscillation probability 
averaged over the various reactor distances. 
But inspite of this averaging effect the KamLAND spectrum data 
reveals an oscillation pattern which enables it to pin down the 
$\Delta m^2_{21}$. However for the KamLAND baseline this pattern 
corresponds to a peak in the survival probability
where the $\theta_{12}$ sensitivity is 
very low. If instead of the peak 
one has a minimum in the survival probability, then the $\theta_{12}$ 
sensitivity can improve dramatically. We show this by simulating
the 3 kTy spectrum for \kl at a  $\Delta m_{21}^2 = 3.5 \times 10^{-5}$ eV$^2$
for which one gets a survival probability 
minimum in KamLAND. 
For this value of $\Delta m_{21}^2$ the spread in $\theta_{\odot}$  
decreases to 25\%, even with the most conservative 6.42\% systematic error. 
We also explore the effect of reducing 
the systematic error to a fictitious value of 2\%. 
This further reduces the error in $\theta_{12}$ to 19\%. 
For the current best-fit value of \dm 
we propose a new \kl like reactor experiment 
with a baseline of $\sim$ 70 km. We show that this 
experiment can observe the minimum in the survival probability and 
therefore the $\theta_{12}$ sensitivity is increased 
by a large amount. For a systematic uncertainty of 2\%, the 
total error in the allowed value of $\tan^2\theta_{12}$ can be 
reduced to about 14\%.

\vskip 10pt
{\bf  Acknowledgment }
The authors would like to thank Raj Gandhi and D.P. Roy for discussions. 
S.C. acknowledges discussions with S.T. Petcov and useful 
correspondences with  Aldo Ianni,  Alessandro Strumia and Francesco Vissani.
S.G. would like to acknowledge a question by 
Yuval Grossman in PASCOS'03 which started this work and D. Indumathi for some 
related comments. The authors express their sincere gratitude to 
Atsuto Suzuki, Fumihiko Suekane and Sandip Pakvasa 
for discussions 
on future systematic errors in KamLAND. 



\end{document}